\documentclass[accepted,creativecommons,sharealike,noncommercial]{eptcs}
\def\event{User Interfaces for Theorem Provers --- UITP 2012}

\usepackage{amsmath}
\usepackage[T1]{fontenc}
\usepackage{graphicx}
\usepackage{listings}
\usepackage{breakurl}
\usepackage{hyperref}

\title{Proof in Context --- Web Editing with Rich, Modeless Contextual Feedback}
\def\titlerunning{Proof in Context}

\author{Carst Tankink
\institute{Institute for Computing and Information Science}
\institute{Radboud University Nijmegen}
\email{carst@cs.ru.nl}}
\def\authorrunning{Carst Tankink}

\def\pa{theorem prover}
\def\uc{\textsc{Update\_client}}
\def\us{\textsc{Update\_server}}

\begin{document}
\maketitle

\begin{abstract}
The Agora system is a prototypical Wiki for formal mathematics: a web-based system for collaborating on formal mathematics, intended to support informal documentation of formal developments. This system requires a reusable proof editor component, both for collaborative editing of documents, and for embedding in the resulting documents.
This paper describes the design of Agora's asynchronous editor, that is generic enough to support different tools working on editor content and providing contextual information, with interactive \pa s being a special, but important, case described in detail for the Coq \pa.
\end{abstract}

\section{Introduction}
The Agora\footnote{\url{http://mws.cs.ru.nl/agora_ui}} system is a prototype for a ``Wiki for Formalized Mathematics'' \cite{CorbineauKaliszyk-2007}: it provides web-based access to repositories of formal documents, allowing authors to write informal descriptions that include snippets of formal text \cite{TankinkLangeUrban-2012}.

Many formal documents are written in an iterative fashion: the author of a formal document writes commands for an interactive \pa, which interprets them to manipulate a ``proof state'': a list of assumptions and a goal that should follow from them. Based on this state, the author then writes new commands for the \pa, until the initial goal is dismissed as proven. Taken together, these commands form a proof \emph{script}, which can be distributed to other users of the \pa. Because the script is meaningless without the proof states, a system that wants to give readers stand-alone access to the proofs should provide the proof states as well, a model which we explored with the Proviola tool~\cite{Tankink+-2010}.

For a Wiki, it is not enough to just offer read-only access to the proofs: one of the main design principles for the first Wiki was that \emph{anyone} can edit \emph{anything}, even if just by a little bit \cite{Cunningham-2003}. Additionally, because formal proofs are similar to computer programs, reader understanding can be improved by allowing the reader to interact with the material in a ``sandbox'': an editor embedded in a document, that includes the material of the document for the reader to play with, for example to redo steps of the proof in a different way, or attempting to apply proven lemmas in (slightly) different situations.
For formal proof, there are two issues barring the way to an accessible editing experience:

\begin{description}
	\item[Verification] The appeal of a Wiki for formal mathematics is that its (formal) content is verified by a proof assistant. Because a proof script rarely stands alone, but builds on other documents in a collection, each change to a document should lead to verifying the documents in the Wiki, with respect to that change, which can take a long time, which the user might not be willing to spend.
        \item[Interaction] Because proof scripts are written interactively, a web-based editor should provide interaction. A standard Wiki editor, on the other hand, is intended for writing text containing simple markup and hyperlinks: tasks done in batch mode: the HTML is rendered after editing, instead of giving feedback to the author while editing.
\end{description}

We leave the first issue for future work: Alama \textit{et al.}  \cite{Alama+-2011} have presented some solutions to this issue on the system-and-tool level --- improving the way \pa s process proofs and using version control mechanisms to manage the impact of change, but we believe that a part of the solution can be contributed by the user interface community: by providing interfaces that allow authors to determine the impact of their changes, and allow them to make changes in several documents before contributing the changeset as a whole: this is not the task of an editor on its own, however, so not addressed in this paper.

This paper describes a solution to the second issue: an editor that supports interaction with the content of the Wiki, an editor that is mature enough to allow authors to edit existing documents, but accessible enough to be included as a sandbox for readers with little exposure to a \pa. 
To lower the threshold, we do not reuse the existing ProofWeb editor \cite{Kaliszyk-2007} or the Matita web editor~\cite{Asperti+-2012}: these editors both use the ``Proof General'' \cite{Aspinall-2000} editing model which requires an extra action by the user, clicking a ``proceed'' button, before the \pa\ executes the commands written. This does not invite users to experiment, and can be, as we show here, entirely avoided.

Instead of using a lock-stepped model, we subscribe to an editor supported by \emph{rich, modeless feedback} \cite{Cooper+-2007}: feedback to user changes that is given through a variety of means such as line highlighting and state windows (the rich part) and which is given as soon as possible, without forcing the user out of an editing ``mode'' (the modeless part). This model is similar to the document-oriented Isabelle/jEdit model described by Wenzel \cite{Wenzel-2011}, but in a Web-based setting. This setting gives rise to a challenge and two advantages:

\begin{itemize}
  \item because the client is served on a web page, while the \pa\ resides on a server, all communication is done through the HTTP protocol, which does not support the server pushing data: the challenge is to have the editor poll the server whether new data is available, without losing its reactivity. Techniques for server-push are being integrated into newer versions of web browsers, but the technique does not have a widespread library support: solutions still need to be hand-crafted, which costs as much effort as the design of the current protocol;
  \item the first advantage is that because Agora's editor is just a web page, displaying information is reduced to adding fragments of HTML to the page. This means that arbitrary server-side tools can work on the document, similar to the Isabelle/jEdit model, and that they can communicate their results as HTML fragments, using the existing framework for asynchronous communication. On the contrary, tools that work in the Isabelle/jEdit model that want to report their results to the user would need to implement this display as a part of the jEdit plugin framework, which requires additional implementation work and a deeper knowledge of the jEdit environment, which could be more difficult for some users;
  \item the second advantage is that Agora is web-based, and tools working with proofs can assume to have access to the Internet, and can use this fact to provide relevant information to the user, for example by showing similar formalizations in different \pa s. If an instance of Agora is started offline, it can still work as an editor, giving access to the repository, but not to features requiring an internet connection.
\end{itemize}

In Section \ref{sec:contexts}, we describe these advantages further, explaining what kind of tools can be attached to the editor, and how they can work with a proof. This section also includes a more in-depth look at what we intend the editor to support. In Sections \ref{sec:models} and \ref{sec:sync}, we describe how we overcome the challenge, by describing the document model as it exists on the server, in the client and in transit (Section \ref{sec:models}) and how these incarnations are synchronized (Section \ref{sec:sync}) to hold the same data after one of the representations gets updated. 
In Section \ref{sec:pa_driver}, we describe an important part of our implementation, a driver for the Coq~\cite{Coq} \pa. Because Coq does not support asynchronous computation, it needs to be ``faked'' by adding, and using, extra information in the data structure, making Section \ref{sec:pa_driver} an example of how tools can enrich the data structure for their own purposes. Section~\ref{sec:conclusions} summarizes and gives a perspective on improving and using the editor.

\section{Proving in Agora: Managing Context}
\label{sec:contexts}

\href{http://mws.cs.ru.nl/agora}{Agora} is a prototype Wiki built upon existing repositories of formal mathematics: users can upload existing developments to the system and then collaborate on these documents in a Web-based system. This collaboration can include further development on the formal content, but the primary workflow of the system is based on writing informal pages describing the development, that can contain dynamic and interactive elements. If a document contains formal content, it is possible to bring up the \pa\ state on-demand for any line of formal proof, and readers should be able to do exercises and experiments directly in the web interface.

Since content in Agora is imported from authors' existing formal development, we assume that there is an offline editor that supports more sophisticated workflows, such as writing a proof consisting of definitions spread over multiple files, and Agora's editor can focus on less involved, ``one-off'' editing tasks, in particular:

\begin{description}
  \item[Edits during description] An author describing a formal proof might discover improvements to this content. Changing the code to address these issues should not require the author to change back to the offline editor and resubmit the formal code: it should be possible to edit the formal text inside the Wiki's environment. 
  \item[Exercises] Several text books, most notably the Software Foundations text book \cite{Pierce+-SF}, use a \pa\ to teach formal techniques in computer science. These text books are self-contained formal developments, that a student can run in a local \pa\ installation. Exercises in this text books take the form of formal proofs, which students need to complete. The benefit of a \pa\ is that a student gets direct feedback to whether or not a proof is correct, and that a teacher has a lighter load in verifying student assignments: because the \pa\ has verified the proofs, they will not contain factual mistakes, and a teacher can focus on improving a student's style. On the other hand, the documentation tools accompanying a \pa\ can be used to mark up a text book for online rendering, typically giving better results than the code highlighting in a \pa's offline editors. To combine the benefits of having an online \pa\ verifying student exercises with those of a text book as a rendered HTML page, it is necessary to supply students with an accessible editor that can be embedded in HTML documents.
  \item[Demonstrations] If an author wants to demonstrate a formalization at a location where no \pa\ is available, having an editor at the same place as the formalization, which can easily load this formalization from the Wiki can be useful for showing applications and alternatives.
\end{description}

These use cases are all covered by existing \pa\ environments on the web, such as ProofWeb, but we can lower the threshold further by making editor modeless and generic. Having a generic editor is especially required in Agora, where we want to easily add new theorem provers, and also want provide contextual feedback to the user, drawn from the Wiki.

\paragraph{Proof Context}
When an author writes a proof in Agora, the proof gets translated into a model, discussed in Section \ref{sec:models}, that allows arbitrary tools to work on it: the tools add extra information to the commands the author writes. Because Agora is a web-based system, the tools reside on a server and can do intensive computations, possibly using the Wiki or the entire Web to provide information. Furthermore, because the editor is already in a client-server model, the results are reported asynchronously, without disturbing the author. We call this model of editing \emph{contextual} as the information displayed depends on the location of the text cursor. It is \emph{rich}, using methods of communication beyond text-only dialogs, and \emph{modeless}, computing and reporting information while the author writes.

\begin{figure}
\begin{center}
\includegraphics[width=\textwidth]{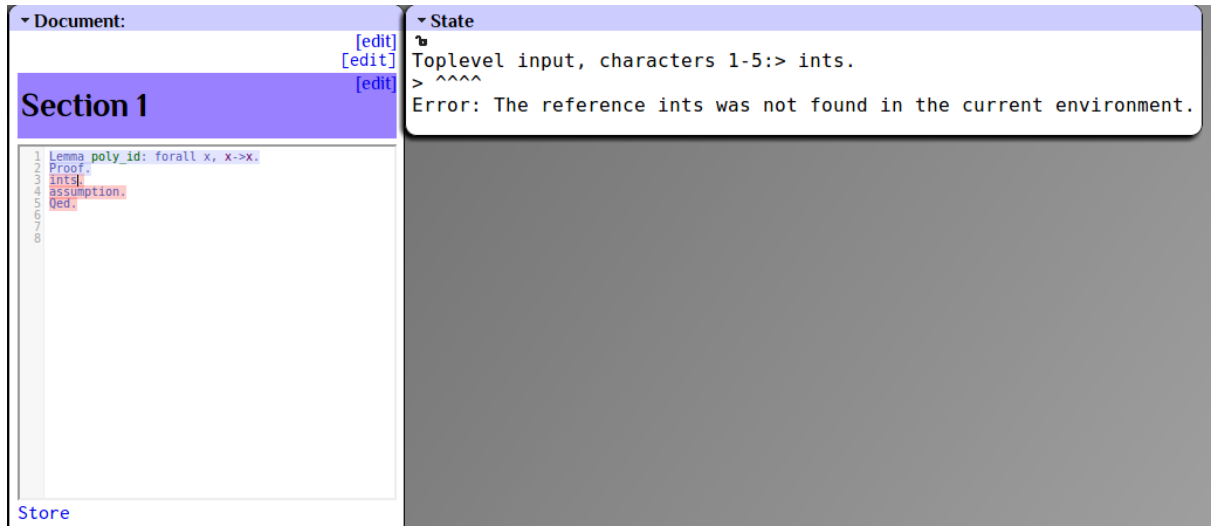}
\end{center}
\caption{Contexts in action}
\label{fig:contexts}
\end{figure}

Figure \ref{fig:contexts} is a screenshot of Agora's editor in action. In this figure, the author has modified a small proof for the Coq \pa. The following contextual information is computed by two server side tools:
\begin{description}
  \item[State] The state window shows an error, reporting that the command under the cursor is incorrect. This response was computed by the \pa.
  \item[Correctness] The first two lines of the proof are correct, the others incorrect: Coq cannot recover from the error on line 3. This is computed by a post-processing step on the \pa\ output, which will be discussed in Section \ref{sec:pa_driver}. 
  \item[Rich type information] In the declaration of the lemma, its name (\texttt{poly\_id}) and the bound variable (\texttt{x}) are coloured differently from the rest of the text. This information is obtained from the so-called ``globalization'' step in Coq's proof process: this step reports what types of identifiers (lemma, variable, \ldots) occur at what locations. This information is computed by the \pa\ during evaluation and the location is reported as a character offset from the beginning of the file. Because the editor adds information to separate commands, a second tool normalizes the information to be a character offset from the beginning of the command.
\end{description}

Other tools can, for example, take the rich type information and evaluate references to lemmas and report them as hyperlinks to the place where they are defined. These hyperlinks can then be displayed as `cheat sheet' window next to the editor.

The list above shows that the \pa\ process is just one of several processes working on a single model. However, the \pa\ has an elevated status: when the author writes a proof, the next command is typically determined by evaluating the state information the tool reports. 
Traditional ``Read-Eval-Print-Loop'' models of \pa\ interaction are based on this driving principle: the author explicitly tells the \pa\ when to proceed on the proof, only proceeding as far as the focus of attention, even if the proof goes on beyond this focus. 

When contextual information is computed based on the proof for the entire proof script, the user should no longer have to instruct the tools to start computing ``on the next line''. Instead, the tools should start computing when new proof text becomes available, reporting their results asynchronously. We give a description of a \pa\ tool that fakes asynchronous updates in Section \ref{sec:pa_driver} as an example of a tool working on the server-side proof, but we first describe how the editor and server represent a proof.

\section{Document representations}
\label{sec:models}

The client and the server share the proof's state in a proof \emph{movie} \cite{Tankink+-2010}, but each represents the model differently. This section describes the different models, including the model used to communicate changes from the server to the client. We describe the server model first: this serves as an introduction to proof movies.

\subsection{Server model}
The server stores the proof as a proof movie: a list of \emph{frames}. These frames contain \emph{cells} that contain the information of the proof: each cell is a piece of information calculated by a tool based on other cells in the frame and, possibly, in previous frames.
Each frame should contain at least a command cell. This cell holds \pa\ commands and comments, calculated by a light-weight parser.

Because we want to link the movie to the client's line-and-character-based model described below, we store the frames not in a list, but in a map from a line-character range to the frame containing the command occurring at that range. We assume that lookup occurs more often than modification, and store the map as a binary tree, indexed by line-character ranges. Because the ranges are contiguous, it is possible to look up a frame by a single text position: given a text position and a key, the position occurs either in the key, letting us return the frame, or it occurs before the key, leading to a descent in the left subtree, or it occurs after the key, leading to a descent in the right subtree.

Given a changed portion of text, we can find the nodes that are affected by this change, and recompute a new tree from this change by taking the affected frames and recomputing new frames from the changed text. After inserting the changed frames, the tree looks like ``preceding frames + changed + following frames'', where the ``preceding'' and ``following'' frames are the old frames that surround the changed portion. Keeping the old frames, especially the preceding ones, allows the \pa\ driver to determine which frames are still valid, improving computation speed, as we describe in more detail in Section \ref{sec:pa_driver}.

\subsection{Client model}
The client model is restricted by the representation of the data structure in HTML: while it is possible to manipulate a data structure as rich as the original movie in JavaScript, it will eventually need to be shown to, and interacted with, by the author in an HTML document. Because of this, and to minimize redundancy of the implementation over several languages, we keep a minimal model in JavaScript, containing only the information we want to show in the client.

In the client, the author needs to have at least the text of the proof, and feedback on how the proof is processed: in terms of the movie, the client needs the commands and proof states for each frame. The commands are placed in some editable text area, the states are requested in response to text cursor movements: following the Isabelle/jEdit model, we take the text cursor position to be indicative of the author's focus of attention, and therefore guiding in what state to show. This heuristic is not completely correct, as is shown by an e-mail thread started by Nipkow in the Isabelle users' mailing list\footnote{\url{https://lists.cam.ac.uk/pipermail/cl-isabelle-users/2012-July/msg00182.html}}: when extending the proof based on the current state, the text cursor moves while typing, causing the system to request the new, possibly non-existing and most likely erroneous, state. On the other hand, when the author has finished typing a command and it is still erroneous, the state can be useful in debugging. More experiments and prototyping is necessary to see how to properly deal with this scenario.

We have a choice in how to represent the text in HTML: we can use a \texttt{textarea} element, a standard HTML DOM element with the \texttt{contenteditable}\footnote{\url{http://www.w3.org/TR/2008/WD-html5-20080610/editing.html\#contenteditable0}} attribute enabled, or an ``off-the-shelf'' editor that is programmable to work with our data.

\paragraph{Textarea} Textarea elements hold plain text and respond to different user editing actions: typing, copying and pasting. Its content cannot be marked up, which makes it not suitable for rich modeless feedback. On the positive side, it does not apply formatting to the text being written, allowing it to be gathered as plain text, which is what the \pa\ expects.

\paragraph{Contenteditable} Using contenteditable on the other hand, causes HTML markup elements to be inserted in non-standard ways, requiring a cleanup of the text before it is processed further. As described by others~\cite{Haverbeeke-2007,Kaliszyk-2007}, this cleanup is non-trivial to implement, and would require effort to be kept synchronized with the models different browsers use. Elements with \texttt{contenteditable} do have the advantage of being just HTML elements, exposed to JavaScript manipulation and CSS styling.

\paragraph{Off-the-shelf editor} A third option is to use a third-party editor component, which allows text to be obtained as plain text, while also providing markup facilities through a programming interface. A feature-rich and easy-to-use representative of this family is CodeMirror\footnote{\url{http://codemirror.net}}, which is also used for several high-profile projects. We choose this editor in order to combine the ease-of-use of a \texttt{textarea} element, while also being able to markup the code the author writes to provide feedback.

An additional benefit of CodeMirror is that it allows a \emph{mark} to be set for a region of text. While the intention of these marks is to apply custom styling to arbitrary regions of text, the fact that each mark is a JavaScript object, and that a JavaScript object can have new attributes added at runtime, allows us to store arbitrary data to regions of text, corresponding to the information found in each frame. This gives us a way to represent the movie in the editor: a frame can be represented by creating a mark on the text in its command cell, and other cells can be added as attributes to this mark. 
When the text cursor is moved, the mark under the cursor can be retrieved, and its data used to render the context.

\subsection{Communication}
For communication, the movie is represented as a list of frames in the ``JavaScript Object Notation'', a textual representation of JavaScript objects, including lists and strings. Instead of sending over the entire frame, a frame is represented by an object, containing the text range its command covers, split in a \texttt{start\_line} and a \texttt{start\_char}, and an \texttt{end\_line} and an \texttt{end\_char}. This suffices, because the client already knows the text content. The rest of the data per frame depends on the computations carried out on the movie, such as those described in Section \ref{sec:pa_driver}: each cell is provided by some tool, with tools providing one or more cells.

\section{Synchronizing the document}
\label{sec:sync}

When the author updates the text in the editor, this should update the server's movie. In turn, this triggers the server-side processes that compute on the new movie. The results of these computations need to be communicated back to the editor. Because the editor should remain reactive when this computation is in progress, and computation takes time, the computation is executed asynchronously and the results reported separately. During computation, the author is free to edit the text, restarting the process. This leads us to distribute the synchronization over two protocols: 

\begin{enumerate}
  \item An \us\ protocol which takes an update of the client text and updates the server model, triggering the start of computation.
  \item An \uc\ protocol which obtains the tool updates to the server model and communicates them to the client.
\end{enumerate}

These two protocols execute asynchronously: \us\ is executed every time the user updates the client model, and \uc\ runs while there is new information to provide to the client, its effect is that the server pushes new information to the client.

From the perspective of the server, there is not much difference between the client's updates to the movie and the updates from other processes, so both protocols might be merged in a single protocol that takes an arbitrary change to a movie and broadcasts it to all other processes. We prioritize the \uc\ protocol for two reasons: first, all computation on the movie is based on the commands, so the other processes are started when new commands come in; second, because the client's only way of communication is through HTTP, which does not support server-side pushes on all browsers, its update protocol requires some extra care.

The figures in the next two sections represent all processes working on the movie as a single, anonymous process $i$. Messages sent to this process should be seen as being sent to all processes.

\subsection{\us: updating the server-side from the client}

\begin{figure}
\begin{center}
	\includegraphics[scale=.9]{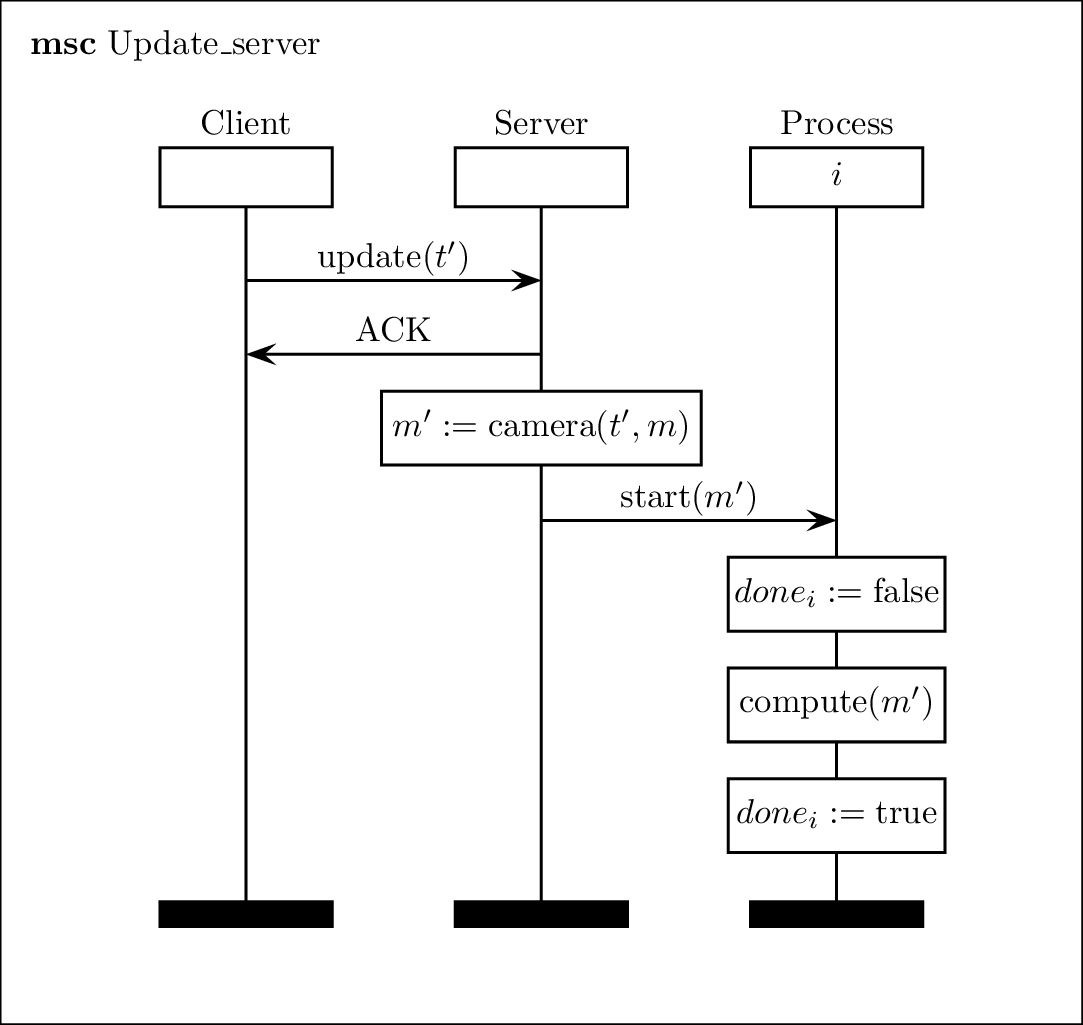}
\end{center}

\caption{The \us\ protocol}
\label{prot:update}
\end{figure}

\us, depicted in Figure \ref{prot:update}, is a straightforward protocol: it is initiated whenever the client's content is updated with new text. It starts by the client sending this text to the server. Immediately upon reception, the server acknowledges this reception, allowing the client to remain responsive. The protocol does not update the client with any information. This information, including what commands are scheduled for computation, is updated by the \uc\ protocol, which should execute as soon as possible after \us. 

The server then creates a movie based on the old movie, $m$, and new text, $t'$, using a \emph{camera} function. This function takes the following steps:
\begin{enumerate}
  \item Get the commands from $m$ as a single text $t$.
  \item Compute the difference $\delta(t, t')$. This difference is a list of ``patch'' operations and should mention at which locations in $t$ changes occur.
  \item Using the locations, find the frames in $m$ that need to be changed, $f_c$, as well as the frames following them: $f_n$. Get the text fragments from the frames $f_c$ and apply the patch to this text, obtaining a new fragment $f'$.
  \item Parse $f'$ into a list of frames.  \label{it:parse}
  \item Insert the new frames and reinsert the frames from $f_n$ to make sure all frames are indexed by their new textual location, clear all tool information stored in the cells of $f_n$, this invalidates all $f_n$, so they will be re-processed. \label{it:update_tree}
\end{enumerate}

The parsing in step \ref{it:parse} transforms the text into a list of commands and comments, based on the \pa's syntax. This is a naive scanner for command terminators described previously for the Proviola tool \cite{Tankink+-2010}, and similar to the \emph{read} phase described by Wenzel \cite{Wenzel-2012}.

Regarding step \ref{it:update_tree}: one could only update the keys of the frames from $f_n$ instead of reinserting the frames. We currently do not use this model for ease of implementation.

After the new tree is computed, all tools subscribed to it are notified, and given a pointer to the new document. These processes can then compute with the new information. After a process has finished computing, it sets a $done$ flag, allowing the server to poll for finished computation in the data retrieval protocol \uc.

\subsection{\uc: getting frame data from the server}
\label{sec:sync-client}

At the end of the \us\ protocol, the server started a number of processes, which work on the content in the movie. During the computation, a process can update frames in the movie with their own data. The goal of the \uc\ protocol is to send this data to the client.

In a normal client-server situation, the server would just push the information to the client, but the client and server communicate over HTTP, which only allows a server to respond to client request. This means we need to mimic server push using JavaScript, using a technique inspired by the \emph{Comet} protocol~\cite{Comet}.

Comet is a technique that emulates server push through a long-standing, asynchronous request: the client requests data from the server, which then waits until it has data to send to the client. The server then responds to the request with this data. This response is delivered through a callback function to the client, meaning the client can continue after requesting data. In the callback function, the client processes the data, and immediately starts a new request.  If there is no data to send, the request is held indefinitely, as long as the implementation allows. Should the request time out, the client can request it again.

A normal Comet implementation assumes that the data on the server side comes in continuously, from a fixed data source. A typical example would be a `ping' command executed by the server, with the results sent to the client. This means that each time the client posts a new request, the server can just read the output from the command, and send it on. 

In our case, the information in the movie comes in bursts: the process can take quite a long while before it has finished computing the information for a single frame, but then quickly fill the next three frames. When this information comes in, the server needs to push it as soon as possible.
 Another difference is the fact that the results of the computations are stored directly in the movie, instead of being streamed directly to the server for it to get when responding to the client. A final piece of the protocol is the fact that processes can be finished; afterwards, they do not produce new data, meaning the server can use the  data it gathered most recently. To allow the server to query tools for new data, we equip each tool with a ``get\_data'' function: a function that, given a frame, reports the data it has computed for that frame. The server can use this function for each frame until the process has stopped. 

We design the \uc\ protocol using the following constraints:
\begin{itemize}
  \item The client initiates the protocol.
  \item The server sends as much data from the processes as is available.
  \item The data is sent as soon as possible.
  \item The server only sends if there is data.
\end{itemize}

  The data is sent as soon as possible to allow the client to respond to the author's editing as soon as possible, giving feedback when it is available, instead of waiting for a potentially long-running computation to finish. This allows the author to react to early problems, without waiting for other problems that might be caused by an earlier error.

The last item is meant to minimize the number of transactions between client and server, to improve the responsiveness of the client. We do not worry that some data might be duplicated among messages, because we expect these messages to be reasonably small.

\begin{figure}[t]
\begin{center}
\includegraphics[scale=.85]{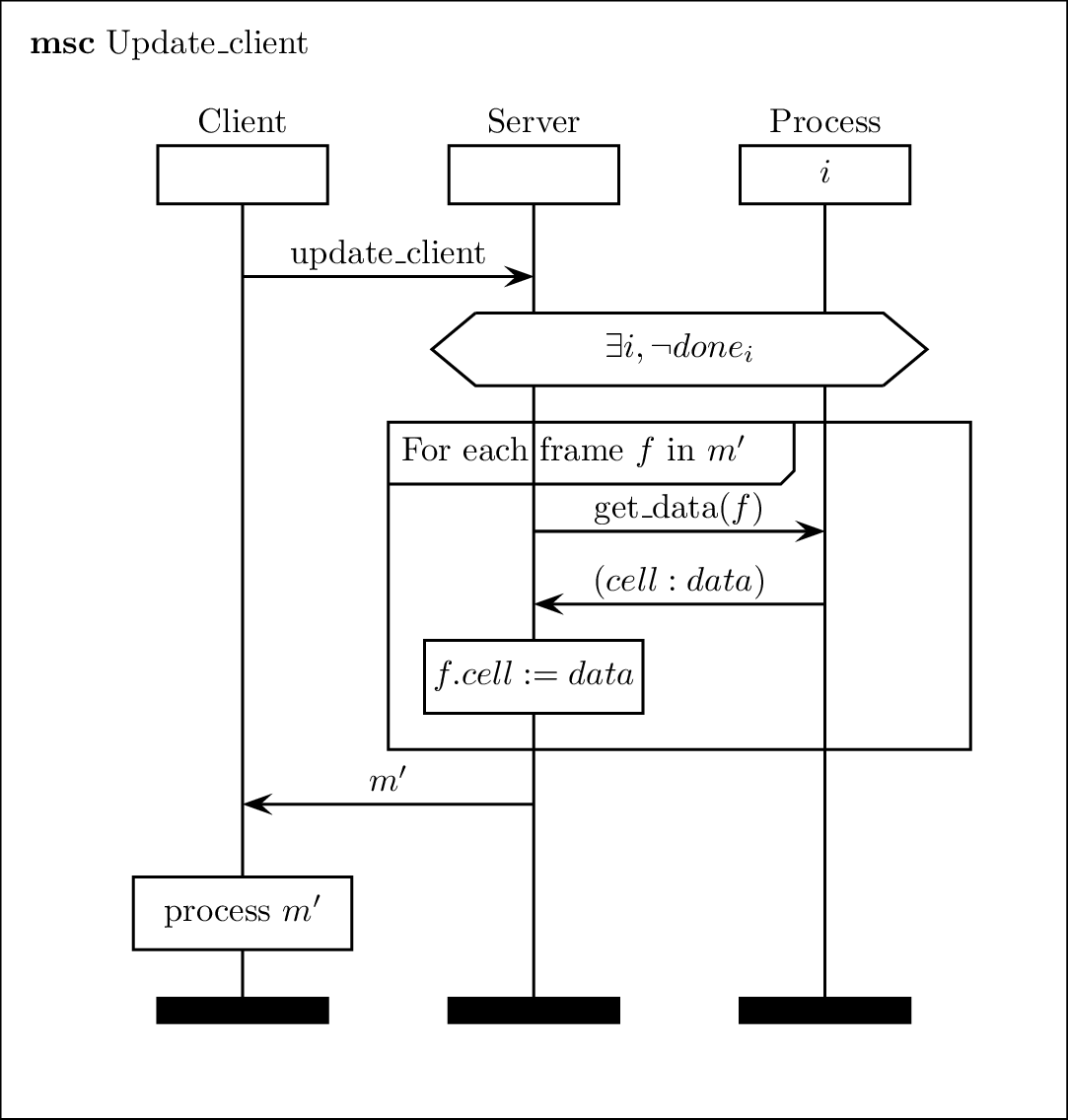}
\end{center}
\label{prot:get_frames}
\caption{The \uc\ protocol}
\end{figure}

The full protocol is shown in Figure \ref{prot:get_frames}. 

This protocol starts the moment the client asks for the movie through the update\_client message. If all processes are finished, the server will hold this request, until at least one of the processes starts again, as a result of the \us\ protocol. When there exist processes which are not finished, the server obtains data from these processes for each frame $f$ in the movie $m'$. The processes return this data as some data, indexed by the cell in which this data belongs. The server then stores the data in the corresponding cell of $f$. After a full pass over the movie in this manner, the new movie is converted to JSON and sent to the client. The client will immediately restart the process and process the data.

This process satisfies our constraints:
\begin{itemize}
  \item The client initiates the protocol through the update\_client message.
  \item By getting data from all processes for each frame, the server sends as much data as is available, assuming the processes report all information computed thus far.
  \item The server updates the movie it sends to the client immediately after a request is made: it does not wait for computations to finish.
  \item The server holds the last request of the client, if there are no processes running. 
\end{itemize}

During the process step, the client uses the data in the movie to add contextual information to the editor text and arrange the information to appear when the text cursor moves, as described in Section \ref{sec:contexts}.

\section{Computing with the documents}
\label{sec:pa_driver}
We have now described the data structures used by client and server and the protocol to synchronize changes in these data structures, but have not yet described the implementation of the processes. There are no requirements on these implementations, apart from reporting data to the server as described in Section \ref{sec:sync-client}. In this section, we describe an implementation of a \pa\ process as an example of the processes that start when the server-side movie is changed.

Processes are connected to changes in the movie following the Observer pattern~\cite{GoF-Observer}: each time the movie updates, the processes are started with the new set of frames. These frames are the only data the processes share with each other and the server. To cache data, the processes are allowed to write to specific cells of each frames. This prevents individual threads of conflicting over the frames and overwriting each other's data, provided the assigned cells are disjoint. 

Having processes produce data in pre-specified cells allows the system to schedule processes that depend on data produced by other processes: each process declares what cells they depend on and what cells they will fill and the scheduler will make sure that once a cell is filled, the interesting parties get notified to start their computation. We will give a concrete example later in this section.

\subsection{A \pa\ driver}
As an example of a more involved process, we describe how we can drive the Coq \pa, based on changes in the movie, by storing enough data in frame cells to make the \pa\ `fake' the behaviour of Isabelle's asynchronous processing~\cite{Wenzel-2012}: while work is underway in making Coq's interaction model asynchronous~\cite{BarrasTassi-2012}, the tool does not yet support this model, and we build a driver around the existing, lock-step, model that allows the server to update the entire movie, and gets states back as efficiently as possible. We do not implement this as a modification for Coq, as the internal interfaces for the \pa\ change more often than the external interfaces, while not allowing more sophisticated state management (unless one digs deeply into the Coq sources).

Before we can describe the driver, we need to explore what frame cells we will have it fill. To do this, we look at how Coq processes a proof script.

\subsubsection{State management and interpretation order}
As explained previously, an interactive \pa\ takes commands and returns responses. This is the first cell our process will fill.
Responses cannot be calculated directly from a single command: instead, a response is computed from a context, consisting of at least a proof state, but which could also include automation hints, notational conventions and other side effects. This means that there is an interpretation order of the frames in a movie: in order to interpret frame $f_2$, the \pa\ first needs to have processed frame $f_1$. In modern models of the \pa, the dependencies between commands are not seen as a list, but as a directed, acyclic graph, the nodes of which can interpreted asynchronously (or lazily), provided all previous nodes have been evaluated. In theory, we support both models: a frame $f$ can hold a list of `dependencies': the frames that need to be evaluated before $f$ gets evaluated. In our implementation, the server assigns these dependencies in a linear fashion, but this function can be replaced by a process doing DAG analysis and requires experimentation with different scheduling strategies.. This would require our driver to depend on a cell holding the dependencies.

When the command of a frame changes, the driver can use the dependencies of the frame to determine which frames it needs to send to the \pa. Naively, it can send all the dependencies from scratch. Less naively, the driver can keep a \pa\ online, and use certain, system-specific commands to 'undo' previous commands, in order to get to the required context. 

\paragraph{Coq's state management}
For Coq, we can obtain best-effort state management by keeping extra administration: for each command, we document which ``state number'' the \pa\ emitted after executing it. The state number reflects the number of commands Coq has executed successfully since it was started and is reported through the program's prompt, provided it is given the \texttt{-emacs} switch. This state number, $n$ can be provided to the \texttt{BackTo} command, causing the system to backtrack to a state $m$, such that $m \leq n$. The caveat for this method is that it is possible to `overshoot' the correct state number: the system is unable to jump back into the proof of a lemma, and will instead skip to the state before the lemma was stated. This process is illustrated in Figures \ref{lst:back_in} and \ref{lst:back_out}, both figures are obtained using Coq 8.4's \texttt{coqtop} tool, started with \texttt{coqtop -emacs}. The responses are replaced by ellipses for the sake of brevity. Additionally, the warning in Figure \ref{lst:back_out} contains non-printing ASCII characters that have been removed before typesetting. 

In Figure \ref{lst:back_in}, the \texttt{BackTo} is given while within the proof of the lemma. This brings the proof state exactly back to the requested number, as evidenced by the last prompt. In Figure \ref{lst:back_out}, on the other hand, the \texttt{BackTo} command is given when the proof is closed. Instead of jumping back into the proof and reverting to state 2, the \pa\ returns to the state before the lemma was stated, throwing a warning.

\begin{figure}
\begin{small}
\begin{lstlisting}
Welcome to Coq 8.4 (October 2012)

<prompt>Coq < 1 || 0 < </prompt>Lemma foo: forall x, x->x.
...
<prompt>foo < 2 |foo| 1 < </prompt>intros.
...
<prompt>foo < 3 |foo| 2 < </prompt>BackTo 2.
...
<prompt>foo < 2 |foo| 1 < </prompt>
\end{lstlisting}
\end{small}
\caption{\texttt{BackTo} within a proof}
\label{lst:back_in}
\end{figure}

\begin{figure}
\begin{small}
\begin{lstlisting}
Welcome to Coq 8.4 (October 2012)

<prompt>Coq < 1 || 0 < </prompt>Lemma foo: forall x, x->x.
...
<prompt>foo < 2 |foo| 1 < </prompt>intros.
...
<prompt>foo < 3 |foo| 2 < </prompt>Admitted.
...
<prompt>foo < 3 |foo| 2 < </prompt>BackTo 2.
Warning: Actually back to state 1.

<prompt>Coq < 1 || 0 < </prompt>
\end{lstlisting}
\end{small}
\caption{\texttt{BackTo} outside of proof}
\label{lst:back_out}
\end{figure}

\subsubsection{Driving Coq}
To make use of the state mechanism in Coq, we use the following system:

\begin{description}
\item[Initialization] Initially, all frames contain a ``reached state'' cell. This cell is initialized to some ``undefined'' value. In practice, any value below 2 can be used, or a value which is not a number at all. Following the order recorded in the ``dependencies'' cell, the commands of the frames are submitted to the \pa. After each frame, the response and the reached state are recorded in the appropriate cells.
\item[On change] When a frame $f$ changes:
\begin{enumerate}
  \item Update its reached state cell to contain the undefined value. Also set the reached state for all following frames, with respect to the order, to undefined: because $f$ sets up a context that the following frames make use of, it is necessary to recompute these frames in the new context. The possibility that the dependency DAG might change as a result of $f$ changing, causing some frames to be ``dangling'', is beyond the scope of this article.
  
  \item For all dependencies of $f$, find the dependency with the highest reached state, $n$.
  \item Send \texttt{BackTo n.} to Coq, record the actually reached state, $m$.
  \item For all frames $f'$ such that the reached state of $f'$ is between $m$ and $n$, send the command of $f'$ to Coq. Coq is now in state $n$.
  \item For each frame that still has a reached state of -1, send that frame's command (with respect to the DAG) recording the reached state after each command.
\end{enumerate} 
\end{description}

This protocol stores extra information in the frames: the cell ``reached state'' contains the state that the \pa\ reported after executing the frame's command. It is used to revert to a relevant state before executing a changed command, as illustrated above. This state can be used for a different purpose: giving feedback about the correctness of a command: whenever a command gets executed successfully, the state counter will increase by one. When a command results in an error, the state is no longer increased. We record this correctness information in its own cell.

In summary, the Coq process only requires command cells, and optionally dependency cells. It fills response, state and correctness cells. The response and correctness cells are used by the client, while the state cell is reused by the process itself, to manage the state of the \pa.

\subsubsection{Expanding to different tools}
We can also use the Coq process to generate globalization information, as described in Section \ref{sec:contexts}. We will not detail this process here. But, when the globalization information is available, a second tool can require the globalization, and generate hyperlinks from this information, in a similar way as Coq's HTML generator, \texttt{coqdoc}. In our framework, this is a process that requires a globalization cell and produces a hyperlink cell. This cell can be used by the client to display the hyperlinks next to the editor or to update the edited text with the hyperlink. The latter does pose a usability question, as the behaviour of a hyperlink click in editable text is ambiguous: it either is meant as a navigation action (following the hyperlink) or as an edit action (placing the cursor to edit the hyperlinked text).

\section{Conclusions and Further Work}
\label{sec:conclusions}
This paper has presented a method of using a \emph{generic} web editor component to communicate with a \pa, by using a shared data structure, without making assumptions about the \pa. In fact, the \pa\ is seen as a generic process working on the data structure, whose results get reported to the client. A prototype implementation can be found at \url{http://mws.cs.ru.nl/agora_ui/}. 

Having this editor available, we can consider the following expansions:

\begin{description}
  \item[Expand to other \pa s] We have currently only implemented communication with the Coq \pa, but it should be possible to adapt other systems to work with the same protocol. In particular, the Isabelle \pa, through its Scala interface, already has a model similar to our movies, so it should be easy to convert between the two models.
  \item[More secondary processes] There are more processes that take a representation of a proof and provide some information about that proof. One example is the proof advisor service for the Mizar \pa~\cite{Urban+-2012}, which provides references to lemmas that will prove the current subgoal.
  \item[Embed in documents] We have not yet embedded our editor in documents in the Wiki, but it would be interesting to post-process, for example, the Software Foundations notes~\cite{Pierce+-SF} to include editors for the example. We imagine these documents to look similar to the interactive tutorial for the CoffeeScript language found at \href{http://autotelicum.github.com/Smooth-CoffeeScript/interactive/interactive-coffeescript.html}{http://autotelicum.github.com/Smooth-CoffeeScript}. 
  \item[Use in Wiki workflow]  Finally, the proof editor has a place in the writing process of Agora: for this, we need to investigate how to best store documents in a Wiki for formal mathematics, and gear the editor to fit in this workflow. In particular, we might want to include tools that provide version control information about the proof documents in the Wiki, as well as information on the impact changing (part of) a proof has on other documents in the Wiki. This might require making the editor aware of versioning, which can also open up new avenues of usability, for example by showing how a proof state evolved while the proof was edited.
\end{description}

Embedding the editor in actual workflows will undoubtedly reveal problems both in implementation and design, but the current prototype shows that it is viable to have a rich, modeless editor for formal proof, that works in a web based setting, lowering the threshold considerably for non-specialists to enter the field.

\bibliographystyle{eptcs}
\bibliography{editing}
\end{document}